\documentclass[final,3p,times]{elsarticle}

\journal{Journal of Process Control}
\usepackage{amssymb}
\usepackage{amsmath}
\usepackage{balance}
\usepackage{mathtools}

\usepackage{graphicx}      
\graphicspath{}
\usepackage{natbib}        
\usepackage{algorithm}
\usepackage{algpseudocode}

\usepackage[T1]{fontenc}
\usepackage{xcolor}
\usepackage{booktabs, makecell, multirow, tabularx,
            threeparttable, tabulary}

\usepackage{setspace}

\usepackage[bottom]{footmisc}
\usepackage{bm}

\newcommand{\abs}[1]{\lvert#1\rvert}

\usepackage{lineno,hyperref}
\modulolinenumbers[5]

\usepackage[capitalise, noabbrev]{cleveref}
\usepackage{caption}
\DeclareCaptionLabelFormat{AppendixTables}{A#2}
\crefname{appendix}{}{}

\begin{document}

\begin{frontmatter}
\title{Meta-Reinforcement Learning for the Tuning of PI Controllers: An Offline Approach\tnoteref{label1}}
\tnotetext[label1]{Please cite using DOI \url{https://doi.org/10.1016/j.jprocont.2022.08.002}}
\author[chbe]{Daniel G. McClement}
\author[math]{Nathan P. Lawrence}
\author[backstrom]{Johan U. Backstr{\"o}m}
\author[math]{Philip D. Loewen$^{\star}$}
\ead{loew@math.ubc.ca}
\author[honeywell]{Michael G. Forbes}
\author[chbe]{R. Bhushan Gopaluni$^{\star}$}
\ead{bhushan.gopaluni@ubc.ca}

\address[chbe]{Department of Chemical and Biological Engineering, University of British Columbia, Vancouver, BC Canada}
\address[math]{Department of Mathematics, University of British Columbia, Vancouver BC, Canada}
\address[backstrom]{Backstrom Systems Engineering Ltd.}
\address[honeywell]{Honeywell Process Solutions, North Vancouver, BC Canada\\
$^{\star}$Authors provided equal supervision.}

\begin{abstract}

Meta-learning is a branch of machine learning which trains neural network models to synthesize a wide variety of data in order to rapidly solve new problems. In process control, many systems have similar and well-understood dynamics, which suggests it is feasible to create a generalizable controller through meta-learning. In this work, we formulate a meta reinforcement learning (meta-RL) control strategy that can be used to tune proportional--integral controllers. Our meta-RL agent has a recurrent structure that accumulates ``context'' to learn a system's dynamics through a hidden state variable in closed-loop. This architecture enables the agent to automatically adapt to changes in the process dynamics. In tests reported here, the meta-RL agent was trained entirely offline on first order plus time delay systems, and produced excellent results on novel systems drawn from the same distribution of process dynamics used for training. A key design element is the ability to leverage model-based information offline during training in simulated environments while maintaining a model-free policy structure for interacting with novel processes where there is uncertainty regarding the true process dynamics. Meta-learning is a promising approach for constructing sample-efficient intelligent controllers.
\end{abstract}


\end{frontmatter}


\section{Introduction}

Reinforcement learning (RL) is a branch of machine learning that formulates a goal-oriented ``policy'' for taking actions in a stochastic environment \citep{sutton1988learning}. This general framework has attracted the interest of the process control community \citep{nian2020review}. For example, one can consider feedback control problems without the need for a process model in this setting. Despite its appeal, an overarching challenge in RL is its need for a significant amount of data to learn a useful policy.\par
\emph{Meta-learning}, or ``learning to learn'', is an active area of research in which the objective is to learn an underlying structure governing a distribution of possible tasks \citep{finn2017model}. In process control applications, meta-learning is appealing because many systems have similar dynamics or a known structure, which suggests training over a distribution could improve the sample efficiency\footnote{How efficient a machine learning model is at learning from data; a high sample efficiency means a model can effectively learn from small amounts of data.} when learning any single task. Moreover, extensive online learning is impractical for training over a large number of systems; by focusing on learning a underlying structure for the tasks, we can more readily adapt to a new system.\par 

This paper proposes a method for improving the \emph{online} sample efficiency of RL agents. Our approach is to train a ``meta'' RL agent \emph{offline} by exposing it to a broad distribution of different dynamics. The agent synthesizes its experience from different environments to quickly learn an optimal policy for its present environment. The training is performed completely offline and the result is a single RL agent that can quickly adapt its policy to a new environment in a model-free fashion. \par
We apply this general method to the industrially-relevant problem of autonomous controller tuning. We show how our trained agent can adaptively fine-tune proportional--integral (PI) controller parameters when the underlying dynamics drift or are not contained in the distribution used for training. We apply the same agent to novel dynamics featuring nonlinearities and different time scales. Moreover, perhaps the most appealing consequence of this method is that it removes the need to accommodate a training algorithm on a system-by-system basis -- for example, through extensive online training or transfer learning, hyperparameter tuning, or system identification -- because the adaptive policy is pre-computed and represented in a single model. \par

\subsection{Contributions}
\label{subsec:contributions}

In this work, we propose the use of meta-reinforcement learning (meta-RL) for process control applications. We create a recurrent neural network (RNN) based policy. The hidden state of the RNN serves as an encoding of the system dynamics, which provides the network with ``context'' for its policy. The controller is trained using a distribution of different processes referred to as ``tasks''. We use this framework to develop a versatile controller which can quickly adapt to effectively control any process from a prescribed distribution of processes rather than a single task.\par
This paper extends \citet{mcclement2021meta} with the following additional contributions:
\begin{itemize}
    \item A simplified and improved meta-RL algorithm: while \citet{mcclement2021meta} required online training, the meta-RL agent in this work is trained entirely offline in advance.
    \item Completely new simulation studies, including industrially-relevant examples dealing with PI controllers and nonlinear dynamics; and
    \item A method of leveraging known, model-based system information offline for the purposes of training, with model-free online deployment.
\end{itemize}
This framework addresses key priorities in industrial process control, particularly:
\begin{itemize}
    \item Initial tuning and commissioning of a PID controller, and 
    \item Adaptive updates of the PID controller as the process changes over time.
    \item Scalable maintenance of PID controllers across many different systems without case-by-case tuning.
\end{itemize}

This paper is organized as follows: In \cref{sec:background} we summarize key concepts from RL and meta-RL; in \cref{sec:algorithm} we describe our algorithm for meta-RL and its practical implementation for process control applications. We demonstrate our approach through numerical examples in \cref{sec:results}, and conclude in \cref{sec:conclusion}.

\subsection{Related work}

We review some related work at the intersection of RL and process control.
For a more thorough overview the reader is referred to the survey papers by~\citet{shin2019ReinforcementLearning, lee2018machine}, or the tutorial-style papers by~\citet{nian2020review, spielberg2019toward}.\par
Some initial studies by~\citet{hoskins1992ProcessControl, kaisare2003simulation, lee2008ValueFunctionbased, lee2010approximate} in the 1990s and 2000s demonstrated the appeal of reinforcement learning and approximate dynamic programming for process control applications. More recently, there has been significant interest in deep RL methods for process control~\citep{noel2014ControlNonlinear, syafiie2011ModelfreeControl, ma2019ContinuousControl, cui2018FactorialKernel, ge2018ApproximateDynamic, pandian2018ControlBioreactor, dogru2021OnlineReinforcement}.\par

\citet{spielberg2019toward} adapted the deep deterministic policy gradient (DDPG) algorithm for setpoint tracking problems in a model-free fashion. Meanwhile,~\citet{wang2018novel} developed a deep RL algorithm based on proximal policy optimization~\citep{schulman2017proximal}. \citet{petsagkourakis2020reinforcement} use transfer learning to adapt a policy developed in simulation to novel systems. Variations of DDPG, such as twin-delayed DDPG (TD3)~\citep{fujimoto2018addressing} or a Monte-Carlo based strategy, have also shown promising results in complex control tasks~\cite{joshi2021ApplicationTwin, yoo2021ReinforcementLearning}. Other approaches to RL-based control utilize a fixed controller structure such as PID~\citep{sedighizadeh2008adaptive, shipman2019ReinforcementLearning, kumar2021diffloop, lakhani2021stability}; some of these are applied to a physical system~\cite{carlucho2017incremental, dogru2022reinforcement, lawrence2022deep}. \par

This present work differs significantly from the approaches mentioned so far. Other approaches to more sample-efficient RL in process control utilize apprenticeship learning, transfer learning, or model-based strategies augmented with deep RL algorithms~\citep{mowbray2021UsingProcess, petsagkourakis2020reinforcement, bao2021DeepReinforcement}. Our method differs in two significant ways. First, the training and deployment process is simplified with our meta-RL agent through its synthesized training over a large distribution of systems. Therefore, only one model needs to be trained, rather than training models on a system-by-system basis. Second, the meta-RL agent in our framework does not rely on precise system identification and only a crude understanding of the process dynamics is required. By training across a distribution of process dynamics, the meta-RL agent learns to control a wide variety of processes with no online or task-specific training required. Although the meta-RL agent is trained in simulation, the key to our approach is that the policy only utilizes process data, and thus achieves efficient model-free control on novel dynamics. A similar concept has been reported in the robotics literature where a robust policy for a single agent is trained offline, leveraging ``privileged'' information about the system dynamics~\citep{lee2020learning}. Most similar to this present work is a paper in the field of robotics where a recurrent PPO policy was trained with randomized dynamics to improve the adaptation from simulated environments to real ones~\citep{DBLP:journals/corr/abs-2006-02402}. \par

\section{Background}
\label{sec:background}

\subsection{Reinforcement learning}
\label{subsec:RL}

In this section, we give a brief overview of deep RL and highlight some popular meta-RL methods. We refer the reader to \citet{nian2020review, spielberg2019toward}, for tutorial overviews of deep RL with applications to process control. We use the standard RL terminology that can be found in \citet{sutton2018reinforcement}. \citet{huisman2021survey} gives a unified survey of deep meta-learning.\par 
The RL framework consists of an \emph{agent} and an \emph{environment}. For each \emph{state} $s_t \in \mathcal{S}$ (the state-space) the agent encounters, it takes some \emph{action} $a_t \in \mathcal{A}$ (the action-space), leading to a new state $s_{t+1}$. The action is chosen according to a conditional probability distribution $\pi$ called a \emph{policy}; we denote this relationship by $a_t \sim \pi(a_t \mid s_t)$. Although the system dynamics are not necessarily known, we assume they can be described as a Markov decision process (MDP) with initial distribution $p(s_0)$ and transition probability $p(s_{t+1} | s_t, a_t)$. A state-space model in control is a special case of an MDP, where the states are the usual (minimal realization) vector that characterizes the system, while the actions are the control inputs. However, the present formulation is more general, as we will demonstrate in later sections. At each time step, a bounded scalar \emph{cost}\footnote{In RL literature, the objective is a maximization problem in terms of a \emph{reward} function. Equivalently, we will formulate a minimization problem in terms of a \emph{cost} function.} $c_t = c(s_t, a_t)$ is evaluated. The cost function describes the desirability of a state-action pair: defining it is a key part of the design process. The overall objective, however, is the expected long-term cost. In terms of a user-specified discount factor $0<\gamma<1$, the optimization problem of interest becomes
\begin{equation}
\begin{aligned}
    &\text{minimize} && J(\psi) = \mathbb{E}_{h \sim p^{\pi_{\psi}}}\left[ \sum_{t=0}^{\infty} \gamma^{t}c(s_t,\pi_{\psi}(a_t \mid s_t)) \right]\\
    &\text{over all} && \psi \in \mathbb{R}^{n}.
\end{aligned}
\label{eq:RLobjective}
\end{equation}
In this formulation, $h$ denotes an infinite-horizon trajectory $h~=~(s_0, a_0, c_0, \ldots, s_N, a_N, c_N, \ldots)$ and the notation $h \sim p^\pi$ indicates that the policy $\pi$ induces a probability distribution $p^\pi$ on the set of such trajectories. Within the space of all possible policies, we optimize over a parameterized subset whose members are denoted $\pi_{\psi}$. We use $\psi$ as a generic term for a vector of parameters: in our application, the individual parameters are weights in a neural network.\par 
Common approaches to solving Problem~(\ref{eq:RLobjective}) involve techniques based on $Q$-learning (value-based methods) and the policy gradient theorem (policy-based methods) \citep{sutton2018reinforcement}, or a combination of both called \emph{actor--critic} methods \citep{konda2000actor}. Closely-related functions to $J$ are the $Q$-function (state-action value function) and value function, respectively:
\begin{align}
    Q(s_t, a_t) &= \mathbb{E}_{h \sim p^{\pi}}\left[ \sum_{k = t}^{\infty} \gamma^{k-t}c(s_k,a_k) \middle| s_t, a_t \right]\label{eq:Qfunc}\\
    V(s_t) &= \mathbb{E}_{h \sim p^{\pi}}\left[ \sum_{k = t}^{\infty} \gamma^{k-t}c(s_k,a_k) \middle| s_t \right].\label{eq:Valuefunc}
\end{align}
The \emph{advantage function} is then $A(s, a) = Q(s, a) - V(s)$. These functions help form the basis for deep RL algorithms, that is, algorithms that use deep neural networks to solve RL tasks. Deep neural networks are a flexible form of function approximators, well-suited for learning complex control laws. Moreover, function approximation methods make RL problems tractable in continuous state and action spaces \citep{lillicrap2015continuous, silver2014deterministic, sutton2000policy}. Without them, discretization of the state and action spaces is necessary, accentuating the ``curse of dimensionality''\footnote{The ``curse of dimensionality'' refers to data sets having exponentially larger ``sample spaces'' as the number of features grows. The larger sample space requires exponentially more training data to learn from, reducing the sample efficiency.}.\par 

A standard approach to solving Problem~(\ref{eq:RLobjective}) uses gradient descent:
\begin{equation}
\psi
\leftarrow \psi - \alpha\nabla J(\psi),
\label{eq:PolicyGradient_Iteration}
\end{equation}
where $\alpha > 0$ is a step-size parameter. 
Analytic expressions for such a gradient exist for both stochastic and deterministic policies \citep{sutton2018reinforcement, silver2014deterministic}. However, in practice, approximations are necessary. Therefore, it is of practical interest to formulate a ``surrogate'' objective that can be used to decrease the true objective given in~(\ref{eq:RLobjective}).\par
Trust region policy optimization (TRPO) is an on-policy method for decreasing $J$ with each policy update \citep{schulman2015trust}. Using the latest policy, whose weights we denote by $\psi_{\text{old}}$, the surrogate objective function is defined as 
\begin{equation}
L_{\psi_{\text{old}}}(\psi) = \mathbb{E}_{h \sim p^{\pi_{\psi_{\text{old}}}}}\left[ \frac{\pi_{\psi}(a \mid s)}{\pi_{\psi_{\text{old}}}(a \mid s)} A_{\psi_{\text{old}}}(s,a) \right] \label{eq:surrogateLoss}
\end{equation}
The surrogate objective function $L_{\psi_{\text{old}}}$ quantifies the advantage of the optimization variable, policy $\pi_\psi$, over the trajectories of the most recent policy, using the old policy $\pi_{\psi_{\text{old}}}$ as an importance sampling estimator. The keys behind the derivation of TRPO are twofold: 1) There exists a non-trivial step-size that will improve the true objective $J$; 2) In order to decrease the true objective, one must place a constraint on the ``difference'' between policies between update iterations. We use the Kullback-Leibler (KL) divergence, defined for generic probability densities $p$ and $q$ by $D_{\text{KL}}( p \parallel q ) = \mathbb{E}_{x \sim p} \left[ \log\left(\frac{p(x)}{q(x)}\right)\right]$. The principal result is that there is constant $C$ such that 
\[\begin{aligned}
&J(\pi) \leq L_{\psi_{\text{old}}}(\psi) + C D_{\text{KL}}^{\text{max}}(\pi_{\psi_{\text{old}}}, \pi)&
\\
&\qquad\text{where}\quad
D_{\text{KL}}^{\text{max}}(\pi, \tilde{\pi}) = \max_{s} D_{\text{KL}}( \pi(\cdot \mid s) \parallel \tilde{\pi}(\cdot \mid s) ),&
\end{aligned}\]
and that minimizing this function over $\psi$ will decrease the true objective $J$ \citep{schulman2015trust}. In practice, TRPO minimizes $L_{\psi_\text{old}}$ subject to a hard constraint on $D_{\text{KL}}^{\text{max}}$ between policy iterates. Regardless of this hard constraint, the optimization problem is solved using natural policy gradients, which requires computing the Hessian of the KL-divergence with respect to the policy parameters. Thus, the main disadvantage of TRPO is its scalability due to its computational burden.\par
Proximal policy optimization (PPO) is a first-order approximation of TRPO \citep{schulman2017proximal}. The main idea behind PPO is to modify the surrogate loss function in \cref{eq:surrogateLoss} such that parameter updates using stochastic gradient descent do not drastically change the policy probability density. The new surrogate objective function is the following:
\begin{equation}
L_{\psi_{\text{old}}}^{\text{PPO}}(\psi) = 
\mathbb{E}_{h \sim p^{\pi_{\psi_{\text{old}}}}}\left[ \max \left\{ \frac{\pi_{\psi}(a \mid s)}{\pi_{\psi_{\text{old}}}(a \mid s)} A_{\psi_{\text{old}}}(s,a),\,
\text{sat}\left( \frac{\pi_{\psi}(a \mid s)}{\pi_{\psi_{\text{old}}}(a \mid s)}; 1, \epsilon \right) A_{\psi_{\text{old}}}(s,a) \right\} \right]\label{eq:PPOsurrogateLoss}
\end{equation}
where $\text{sat}(u; 1, \epsilon) = u$ if $-\epsilon < u - 1 <\epsilon$ and $\text{sat}(u; 1, \epsilon) = 1 + \epsilon \frac{u}{\abs{u}}$ otherwise. Despite being somewhat complicated, the intuition for \cref{eq:PPOsurrogateLoss} is understood through cases inside the `max' functions: when $A$ is positive, the term inside the expectation becomes $\max\left( \frac{\pi_{\psi}(a \mid s)}{\pi_{\psi_{\text{old}}}(a \mid s)}, 1 - \epsilon \right) A_{\psi_{\text{old}}}(s,a)$, which puts a limit on how much the objective can decrease; the case when $A$ is negative is similar. Either way, the term inside the expectation can only decrease by making actions more or less likely, depending on if the advantage is positive or negative, respectively. Moreover, the saturation limits how much the new policy can deviate from the old one. Trajectories with $\pi_{\text{old}}$ are used to approximate $A_{\psi_{\text{old}}}$, which is then used to approximate and optimize \cref{eq:PPOsurrogateLoss} using gradient descent.

\subsection{Meta reinforcement learning}

\begin{figure}
\begin{center}
    \includegraphics[width=0.75\linewidth]{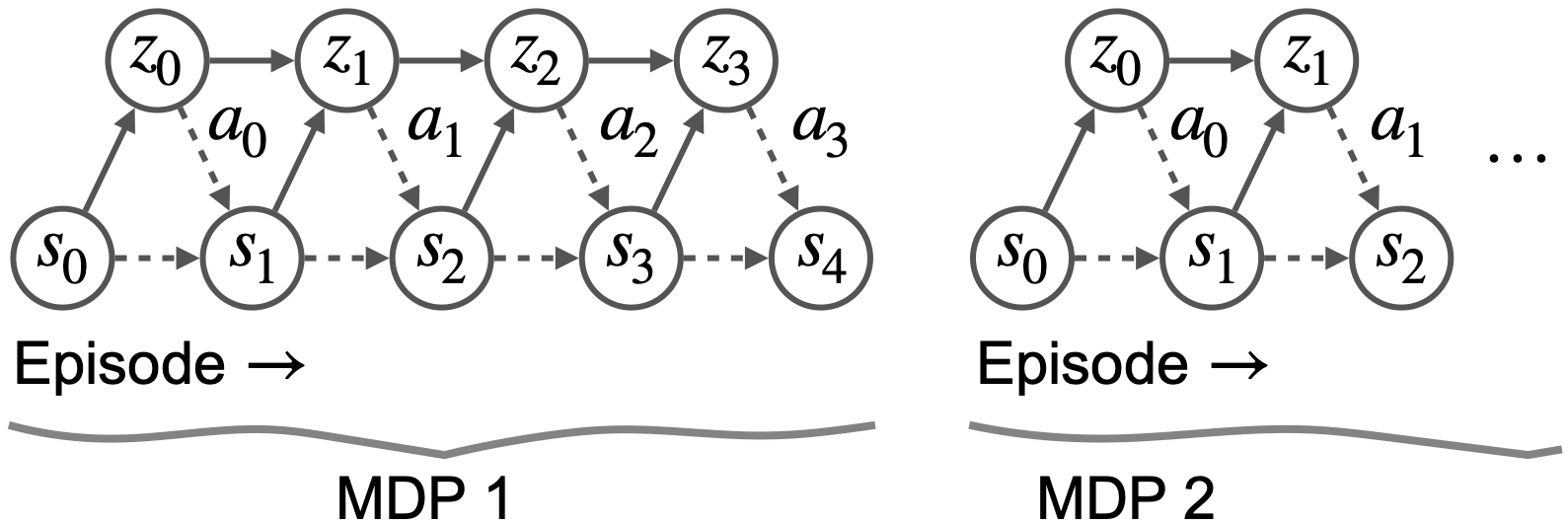}
    \caption{A diagram of the meta-RL agent's interactions with the task distribution $p_{\text{meta}}$.}
    \label{fig:RL2}
\end{center}
\end{figure}

While the algorithms mentioned above can achieve impressive results in a wide range of domains, they are designed to be applied to a single MDP. In contrast, meta-RL aims to generalize agents to a distribution of MDPs. Formally, a single MDP can be characterized by a tuple $\mathcal{T} = (\mathcal{S}, \mathcal{A}, p, c, \gamma)$; in contrast, meta-RL tackles an optimization problem over a distribution $p_{\text{meta}}(\mathcal{T})$ of MDPs. Therefore, in the meta-RL terminology, a ``task'' is simply all the components comprising a single RL problem. The problem of interest in the meta-RL setting is a generalization of the standard RL objective in Problem~(\ref{eq:RLobjective}) \citep{huisman2021survey}:
\begin{equation}
\begin{aligned}
    &\text{minimize} && J_{\text{meta}}(\bm{\Psi}) = \mathbb{E}_{\mathcal{T} \sim p_{\text{meta}}(\mathcal{T})} \left[J(\psi^{*}(\mathcal{T},\bm{\Psi}))\right]\\
    &\text{over all} && \bm{\Psi} \in \mathbb{R}^{n}.
\end{aligned}
\label{eq:Metaobjective}
\end{equation}
Crucially, in the context of process control, meta-RL does \emph{not} aim to find a single controller that performs well across different plants. Note that $\psi^{*}$ in \cref{eq:Metaobjective} is the optimal weight vector in~(\ref{eq:RLobjective}) as a function of a sampled MDP $\mathcal{T}$ and the meta-weights $\bm{\Psi}$. Meta-RL agents aims to simultaneously learn the underlying structure characterizing different plants and the corresponding optimal control strategy under its cost function. The practical benefit is that this enables RL agents to quickly adapt to novel environments.\par

There are two components to meta-learning algorithms: the models (e.g., actor--critic networks) that solve a given task, and a set of meta-parameters that learn how to update the model \citep{bengio2013optimization, andrychowicz2016learning}. Due to the shared structure among tasks in process control applications, we are interested in \emph{context-based} meta-RL methods \citep{duan2016rl, rakelly2019efficient, wang2016learning}. These approaches learn a latent representation of each task, enabling the agent to simultaneously learn the context and the policy for a given task.\par
Our method is similar to \citet{duan2016rl}. We treat the problem in line~(\ref{eq:Metaobjective}) as a single RL problem. For each MDP $\mathcal{T} \sim p_{\text{meta}}(\mathcal{T})$, the meta-RL agent has a maximum number of time steps, $T$, to interact with the environment, called an \emph{episode}. As each episode progresses, the RL agent has an internal hidden state $z_t$ which evolves with each time step through the MDP based on the RL states the agent observes: $z_t = f_{\bm{\Psi}}(z_{t-1},s_t)$. The RL agent conditions its actions on both $s_t$ and $z_t$. An illustration of this concept is shown in \cref{fig:RL2}. Therefore, the purpose of the meta-parameters $\bm{\Psi}$ is to quickly adapt a control policy for an MDP $\mathcal{T} \sim p_{\text{meta}}(\mathcal{T})$ by solving for a suitable set of MDP-specific parameters encoded by $z_t$. This is why this approach is described as meta-RL; rather than training a reinforcement learning agent to control a process, we are training a meta-reinforcement learning agent to find a suitable set of parameters for a reinforcement learning agent which can control a process. The advantage of training a meta-RL agent is that the final model is capable of controlling every MDP across the task distribution $p_{\text{meta}}$ whereas a regular RL agent could only be optimized for a fixed task $\mathcal{T}$.\par
Clearly, the key component of the above framework is the hidden state. This is generated with a recurrent neural network (RNN), which we briefly describe in a simplified form. An RNN is a special neural network structure for processing sequential data. Its basic form \citep{goodfellow2016deep} is shown below:
\begin{align}
    z_t &= f \left( W z_{t-1} + U x_t + b \right) 
    \label{eq:rnn1}
    \\
    o_t &= V z_t + c.
    \label{eq:rnn equation}
\end{align}
Here $W, U, V, b, c$ are trainable weights, while $x_t$ is some input to the network and $o_t$ is the output. $f$ is a nonlinear function. An RNN can be thought of as a nonlinear state-space system that is optimized for some objective. The characteristic feature of any type of RNN is the hidden state, which evolves alongside sequential input data. The simple RNN formulation in Equations \ref{eq:rnn1} and \ref{eq:rnn equation} is prone to vanishing or exploding gradients. In practice, we mitigate these problems by using a more sophisticated form of recurrent layer called the gated recurrent unit (GRU). GRUs use trainable information gates to control the updates to a layer's hidden state which help avoid vanishing gradients, the reader is referred to \citep{cho2014learning} for further information on the GRU architecture.

\section{Meta-RL for process control}
\label{sec:algorithm}

We apply the meta-RL framework to the problem of tuning proportional--integral (PI) controllers. The formulation can be applied to any fixed-structure controller, but due to their prevalence, we focus on PI controllers as a practical illustration.

\subsection{Tasks, states, actions, costs}

The systems of interest are first-order plus time delay (FOPTD): their transfer functions have the form
\begin{equation}
    G(s) = \frac{K}{\tau s + 1} e^{-\theta s},
    \label{eq:foptd}
\end{equation}
where $K$ is the process gain, $\tau$ is the time constant, $\theta$ is the time delay, and $s$ is the Laplace variable (not to be confused with $s_t$, the RL state at time step $t$). Such models are often good low-order approximations for the purposes of PI tuning \citep{skogestad2003simple}. The formulation in continuous time is tidy, but in practice we of course discretize \cref{eq:foptd}.\par
A PI controller has the form
\begin{equation}
    C(s) = K_c \left (1 + \frac{1}{\tau_I s} \right),
    \label{eq:PI standard form}
\end{equation}
where $K_c$ and $\tau_I$ are constant tuning parameters.
In our numerical work, we used $k_p=K_c$ and $k_i=K_c/\tau_I$ 
instead of $K_c$ and $\tau_I$

in the RL state to improve the numerical stability of the computations\footnote{The inverse relationship between $\tau_I$ and the controller output can cause instability early in offline training, if a poorly trained meta-RL model sets $\tau_I\approx0$. No similar stability concerns arise when using $k_i$.}.\par

Prior work on RL for PI tuning suggests an update scheme of the form \citep{lawrence2022deep}:
\begin{align}
    [k_p, k_i] &\leftarrow [k_p, k_i] + \alpha \nabla J([k_p, k_i])\\
    &=  [k_p, k_i] + \Delta [k_p, k_i]
\end{align}
where the RL policy is directly parameterized as a PI controller. Therefore, in the meta-RL context, we take the actions to be changes to the PI parameters $\Delta [k_p, k_i]$.\par

For simplicity, the MDP state ($s_t$) used by the RL agent to select its actions (updates to the PI parameters) is based on the standard form of the PI controller. In practice, different flavours of fixed-structure controllers can be used, including PI controllers in velocity form and full PID controllers. The MDP state at time step $t$ contains the PI parameters, the proportional setpoint error and the integrated setpoint error from the beginning of the episode, $t_0$: 
\begin{equation}
    s_t = \left[k_p, k_i, e_t, \int_{t_{0}}^{t} e_{\tau} d\tau \right].
    \label{eq:state definition}
\end{equation}

The RL agent is trained to minimize its discounted future cost interacting with different tasks. The cost function used to train the meta-RL agent is the squared error from a target trajectory, shown in \cref{eq:cost}. The target trajectory is calculated by applying a first order filter to the setpoint signal\footnote{The filtered setpoint signal is only used to calculate the meta-RL agent's cost function. The PI controller itself does not use this filtered setpoint signal; the controller uses the unfiltered setpoint signal when calculating control actions.}. The time constant of this filter is set to the desired closed-loop time constant, $\tau_{cl}$. A target closed-loop time constant of $2\tau$ is chosen for robustness and smooth control action, though other choices for $\tau_{cl}$ could be made by a control practitioner, such as setting $\tau_{cl}$ to the process dead time \cite{skogestad2003simple}. An $L^1$ regularization penalty $\beta > 0$ on the agent's actions is also added to the cost function to encourage sparsity in the meta-RL agent's output and help the tuning algorithm converge to a constant set of PI parameters (rather than acting as a nonlinear feedback controller and constantly changing the controller parameters in response to the current state of the system).

\begin{equation}
    c_t = (y_{\text{desired},t} - y_t)^2 + \beta_1 |\Delta k_p| + \beta_2 |\Delta k_i|,
    \label{eq:cost}
\end{equation}

\begin{equation}
    Y_{\text{desired}}(s) = \frac{y_{sp}}{2\tau s + 1}e^{-\theta s}
    \label{eq:Ydesired}
\end{equation}

Comparing the RL state definition to the RL cost definition, we see similar trajectories through different MDPs will receive very different costs depending on the underlying system dynamics in the particular tasks being controlled. In order for the meta-RL agent to perform well on a new task, it needs to perform implicit system identification to generate an internal representation of the system dynamics.

The advantages of this meta-RL scheme for PI tuning are summarized as follows:
\begin{itemize}
    \item Tuning is performed in closed-loop and without explicit system identification.
    \item Tuning is performed automatically even as the underlying system changes.
    \item The agent can be deployed on new systems within the task distribution $p_{\text{meta}}$ without any online training. Further, as shown in \cref{subsec:tank-results}, nearly any system can be modified to be ``in-distribution'' in this sense.
    \item The meta-RL agent is a single model that is trained once, offline, so there is no need to specify hyperparameters on a task-by-task basis.
    \item The meta-RL agent's cost function is conditioned on the process dynamics and will produce consistent closed-loop control behaviour on different systems.
\end{itemize}
This approach is not limited to PI tuning. It can also be applied to other scenarios where the model \emph{structure} is known. The agent then learns to behave near-optimally inside each task in the training distribution, bypassing the need to identify model parameters and only train on that instance of the dynamics.

\subsection{RL agent structure}
\label{subsec:agent}

\begin{figure}
\begin{center}
    \includegraphics[width=0.5\linewidth]{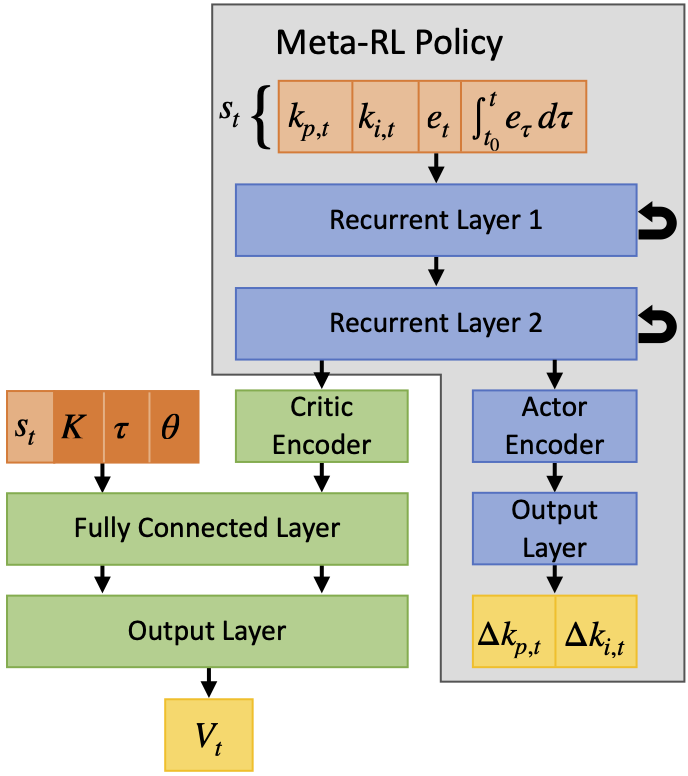}
    \caption{The structure of the RL agent. The control policy used online is shown in the grey box while the critic used during offline training is shown in green.}
    \label{fig:controller-structure}
\end{center}
\end{figure}

The structure of the meta-RL agent is shown in \cref{fig:controller-structure}. The grey box shows the ``actor'', i.e., the part of the agent used online for controller tuning. Through interacting with a system and observing the RL states at each time step, the agent's recurrent layers create an embedding (hidden state) which encodes information needed to tune the PI parameters, including information about the system dynamics and the uncertainty associated with this information. These embeddings essentially represent process-specific RL parameters which are updated as the meta-RL agent's knowledge of the process dynamics changes. Two fully connected layers use these embeddings to recommend adjustments to the controller's PI parameters. The inclusion of recurrent layers is essential for the meta-RL agent's performance. Having a hidden state carried between time steps equips the agent with memory and enables the agent to learn a representation of the process dynamics. A traditional feedforward RL network would be unable to differentiate between different tasks and would perform significantly worse. This concept is demonstrated in \citet{mcclement2021meta}.\par

Outside of the grey box are additional parts of the meta-RL agent which are only used during offline training. The ``critic'' (shown in green) is trained to calculate the value (an estimate of the agent's discounted future cost in the current MDP given the current RL state). This value function is used to train the meta-RL actor through gradient descent using \cref{eq:PPOsurrogateLoss}.\par

A unique strategy we use to improve the training efficiency of the meta-RL agent is to give the critic network access to ``privileged information'', defined as any additional information outside the RL state and denoted by $\zeta$. In addition to the RL state, the critic conditions its estimates of the value function on the true process parameters ($K$, $\tau$, and $\theta$), as well as the deep hidden state\footnote{The deep hidden state is the hidden state of the second (i.e. ``deeper'') recurrent layer in the meta-RL agent.} of the actor. Knowledge of a task's process dynamics, as well as knowledge of the actor's internal representation of the process dynamics through its hidden state, allows the controller to more accurately estimate the value function, which improves the quality of the surrogate objective function used to train the actor. Equipping the critic with this information also allows it to operate as a simpler feedforward neural network rather than a recurrent network like the actor.\par

The privileged information given to the critic network may at first appear to conflict with the advantages of the proposed meta-RL tuning method, since the critic requires the true system parameters and much simpler tuning methods for PI controllers exist if such information is known. However, this information is only required during \emph{offline} training. The meta-RL agent is trained on simulated systems with known process dynamics, but the end result of this training procedure is a meta-RL agent that can be used to tune PI parameters for a real process \emph{online} with no task-specific training or knowledge of the process dynamics. The portion of the meta-RL agent operating online contained in the grey box only requires RL state information -- process data -- at each time step.

\subsection{Training algorithm}
The meta-RL agent is trained by uniformly sampling $K$, $\tau$, and $\theta$ to create a FOPTD system and initializing a PI controller with $K_c=0.05$ and $\tau_I=1.0$. These initial PI tunings were selected because they result in a very slow control response for any possible system from the meta-RL training distribution. The ultimate performance of the PI controller can then be attributed to the meta-RL agent's tuning rather than a good initialization of the controller's parameters. The state of the system is randomly initialized near zero and the setpoint is switched between $1$ and $-1$ every $11$ units of time. The meta-RL agent has no inherent time scale and so we keep the units of time general to highlight the applicability of the proposed PI tuning algorithm to both fast and slow processes (allowing time constants whose orders range from milliseconds to hours).\par

\cref{tab:foptd distribution} shows the ranges from which the FOPTD model parameters are uniformly sampled  during training. In \cref{subsec:tank-results} we demonstrate how data augmentation extends the applicability of training across this range of parameters.\par

There are two main limitations to the size of the task distribution the meta-RL agent can effectively be trained across. First, neural network training works best when the features of interest have a consistent scale. However, for different systems, suitable $k_p$ and $k_i$ parameters can vary by orders of magnitude. It becomes very difficult to train a neural network to effectively process inputs with significantly varying magnitudes ($k_p$ and $k_i$ are part of the RL state) as well as produce outputs which vary by orders of magnitude ($\Delta k_p$ and $\Delta k_i$ are the RL actions). Second, the time scale of the distribution of systems must be reasonably bounded so there exists a sampling time\footnote{The sampling time referenced in this work is the sampling time for updates to the RL state (which is also the time increment between updates to the controller gains). This is \emph{not} the same as the controller sampling time used to update the control action.} for the meta-RL agent to use which is appropriate for every system it interacts with. A large MDP time step on systems with fast dynamics will not allow the meta-RL agent to effectively learn the process dynamics. The transient response to any setpoint change or disturbance would occur between time steps and not be visible to the neural network. On the other hand, a small sampling time on systems with slow dynamics will cause transient system responses to stretch across many time steps. Recurrent neural networks struggle to learn relationships in data occurring over very long sequences, so the ability for the network to identify systems with slow dynamics is reduced if the time step is too small.\par

\begin{table}[!ht]
    \begin{center}
    \label{tab:foptd distribution}
    \begin{tabular}{l l l l}
      \toprule 
      Model Parameter & $K$ & $\tau$ &  $\frac{\theta}{\tau}$\\
      \midrule 
      Minimum & 0.25 & 0.25 & 0\\
      Maximum & 1.0 & 1.0 & 1.0\\
      \bottomrule 
    \end{tabular}
    \end{center}
    \caption{The range of model parameters used to train the meta-RL agent.}
    \label{table:trainingranges}
\end{table}

Algorithm 1 shows the procedure used to train the meta-RL agent. Simulations and model training were performed in Python 3 using the PyTorch machine learning library \cite{pytorch}. We started with the PPO algorithm as implemented in Open AI's ``Spinning Up''  \cite{SpinningUp2018} and modified it to accommodate a recurrent neural network and a distribution of control tasks. Hyperparameters are listed in \cref{tab:network-hyperparameters}.

{
\begin{minipage}[]{\linewidth}
\linespread{1}
\footnotesize
\begin{algorithm}[H]
\caption{Meta-RL Controller Training \newline \textit{Adapted from the documentation of OpenAI's PPO } \cite{SpinningUp2018}}
\textbf{Input:} Initial meta-policy parameters $\Psi_0$, initial value function parameters $\phi_0$ 
\begin{algorithmic}[1]
\For {each training episode}
    \State Sample a batch of $n$ tasks $\mathcal{T}_i \sim p_{\text{meta}}(\mathcal{T})$, $i\in\{1,\ldots,n\}$
    \State Initialize a buffer to hold state transition data $\mathcal{D}_k$
    \For {each $\mathcal{T}_i$}
        \State Collect a trajectory $h$ using the current meta-policy $\pi_{\Psi}$ on task $\mathcal{T}_i$
        \State Store $h$ in $\mathcal{D}_k$
    \EndFor
    \State Compute advantage estimates $\widehat{A}_t$ using generalized advantage estimation \cite{schulman2018highdimensional} and the current value function~$V_{\phi}$.
    \State Divide trajectories into sequences of the desired length, $T$, for backpropagation through time.
    \State Update the policy by minimizing the PPO-Clip objective using gradient descent:
    \begin{equation}
    \displaystyle\Psi_{k+1} = \mathrm{arg}\text{ }\underset{\Psi}{\mathrm{min}} \frac{1}{|\mathcal{D}_k|T}\sum_{h \in \mathcal{D}_k}\sum_{t=0}^T \mathrm{max}\left\{ \frac{\pi_{\Psi}(a_t|s_t)}{\pi_{\Psi_k}(a_t|s_t)}\widehat{A}_t, \text{sat}\left( \frac{\pi_{\Psi}(a_t|s_t)}{\pi_{\Psi_k}(a_t|s_t)}; 1, \epsilon \right) \widehat{A}_t \right\}
    \end{equation}
    \State Update the value function to estimate the cost-to-go of an episode using gradient descent:
    $$\displaystyle\phi_{k+1} = \mathrm{arg}\text{ }\underset{\phi}{\mathrm{min}}\frac{1}{|\mathcal{D}_k|T}\sum_{h \in \mathcal{D}_k}\sum_{t=0}^T(V_{\phi}(s_t,\zeta_t)-\widehat{R}_t)^2$$
\EndFor
\end{algorithmic}
\end{algorithm}
\end{minipage}
}

\section{Experimental results}
\label{sec:results}

\subsection{Asymptotic performance of the meta-RL tuning algorithm}
\label{subsec:asymptotic-performance}

\begin{figure}
    \centering
    \includegraphics[width=\linewidth]{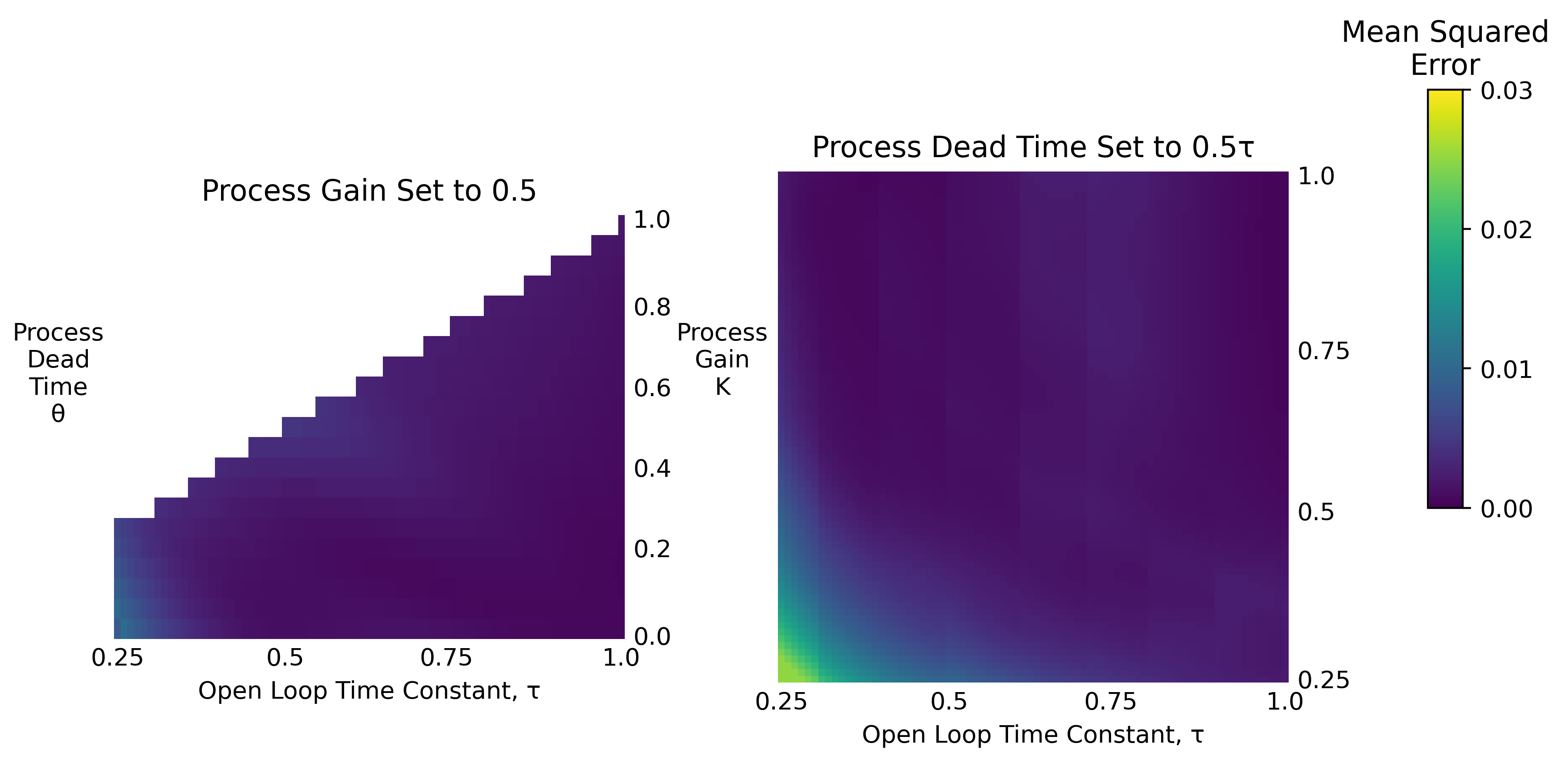}
    \caption{Tracking errors for various FOPTD systems under meta-RL supervision.
    Each point corresponds to a unique combination of $(K_c,\tau,\theta)$.
    Its color shows the corresponding closed-loop system's mean-squared deviation
    from the target response after the meta-RL agent has acted long enough for
    the PI parameters to stabilize.}
    \label{fig:error-heatmap}
\end{figure}

\begin{figure}
    \centering
    \includegraphics[width=\linewidth]{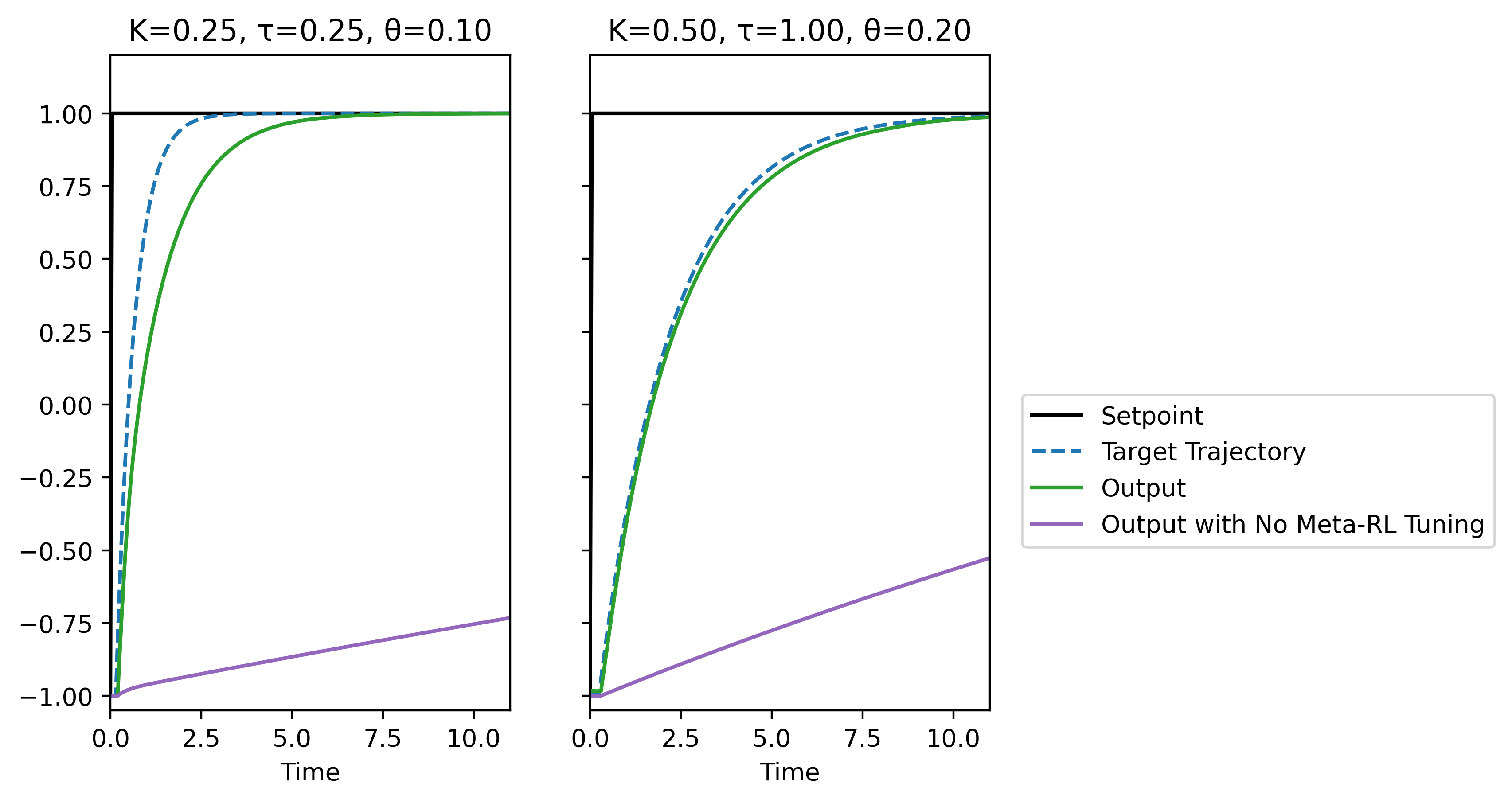}
    \caption{System output trajectories for a setpoint change from $-1$ to $1$ using the meta-RL algorithm's PI tunings compared to the target trajectories. The worst-case (left) and best-case (right) are selected from the heatmaps in \cref{fig:error-heatmap}. A trajectory using the initial PI tunings is also shown for comparison.}
    \label{fig:error trajectory}
\end{figure}

\cref{fig:error-heatmap} depicts the asymptotic performance of the meta-RL tuning method.
The intervals of $K$, $\tau$, and $\theta/\tau$ in \cref{table:trainingranges} define a 3D box
in which each point corresponds to a different FOPTD system. After using the meta-RL agent to
generate a PI controller for every such system, we could apply a setpoint step from $-1$ to $1$,
observe the closed-loop response (see \cref{fig:error trajectory}),
and compute its mean-squared deviation from the target trajectory in \cref{eq:Ydesired}.
The results could, in principle, be used to produce a solid 3D heatmap. 
\cref{fig:error-heatmap} shows two heatmaps sliced from this solid.
In the left subplot, 
$K$ is held constant at $0.5$, while $\tau$ and $\theta$ vary on the horizontal and vertical axes.
On the right, 
$\frac{\theta}{\tau}$ is held constant at $0.5$, while $\tau$ and $K$ vary on the horizontal and vertical axes.
The dark color dominating both images corresponds to a very small mean-squared error.
Lighter shades in the right subplot indicate larger MSE values for systems where both $K_c$ and $\tau$ are small.
The system for which the MSE is largest is labelled with a red dot.
Overall, every system in the given region of parameter space is well-controlled by the meta-RL tuning algorithm.

\cref{fig:error trajectory} depicts the performance in the worst-case and best-case scenarios based on target trajectory tracking performance selected from \cref{fig:error-heatmap}. In the best-case, the mean squared error between the meta-RL's control trajectory and the target trajectory is 0.0004 while in the worst-case the mean squared error is 0.0300. Even in the worst-case scenario, the meta-RL algorithm's PI tunings provide desirable control performance. \cref{tab:pituning} compares the meta-RL agent's PI tunings in the worst-case and best-case scenario to the PI tunings calculated using the improved SIMC PI tuning method~\cite{improved_simc}, which provides near-optimal tunings for FOPTD systems. In the best-case (which from Figure 3 we see is similar to the performance across most of the task distribution), there is a 2.99\% difference between the meta-RL PI tunings and the SIMC tunings.\par

\begin{table}[tb]
\caption{Comparison between the meta-RL agent's PI tunings and those calculated using the SIMC method \cite{improved_simc}.}
  \small
\centering
  \begin{tabularx}{\linewidth}{X  XX  XX} 
    \toprule
     &
      \multicolumn{2}{c}{$K_c$} & 
      \multicolumn{2}{c}{$\tau_I$} \\
      \cmidrule(rl){2-3}
      \cmidrule(rl){4-5}
     System & {Meta-RL} & {SIMC} & {Meta-RL} & {SIMC} \\
          \midrule
    Best-case ($k=0.5$, $\tau=1.0$, $\theta=0.2$) & 0.876 & 0.909 & 1.042 & 1.067 \\
    Worst-case ($k=0.25$, $\tau=0.25$, $\theta=0.1$) & 1.227 & 1.667 & 0.367 & 0.283\\
    \bottomrule
  \end{tabularx}
\label{tab:pituning}
\end{table}

\subsection{Online sample efficiency of the meta-RL tuning algorithm}
\label{subsec:time}

\begin{figure}
    \centering
    \includegraphics[width=\linewidth]{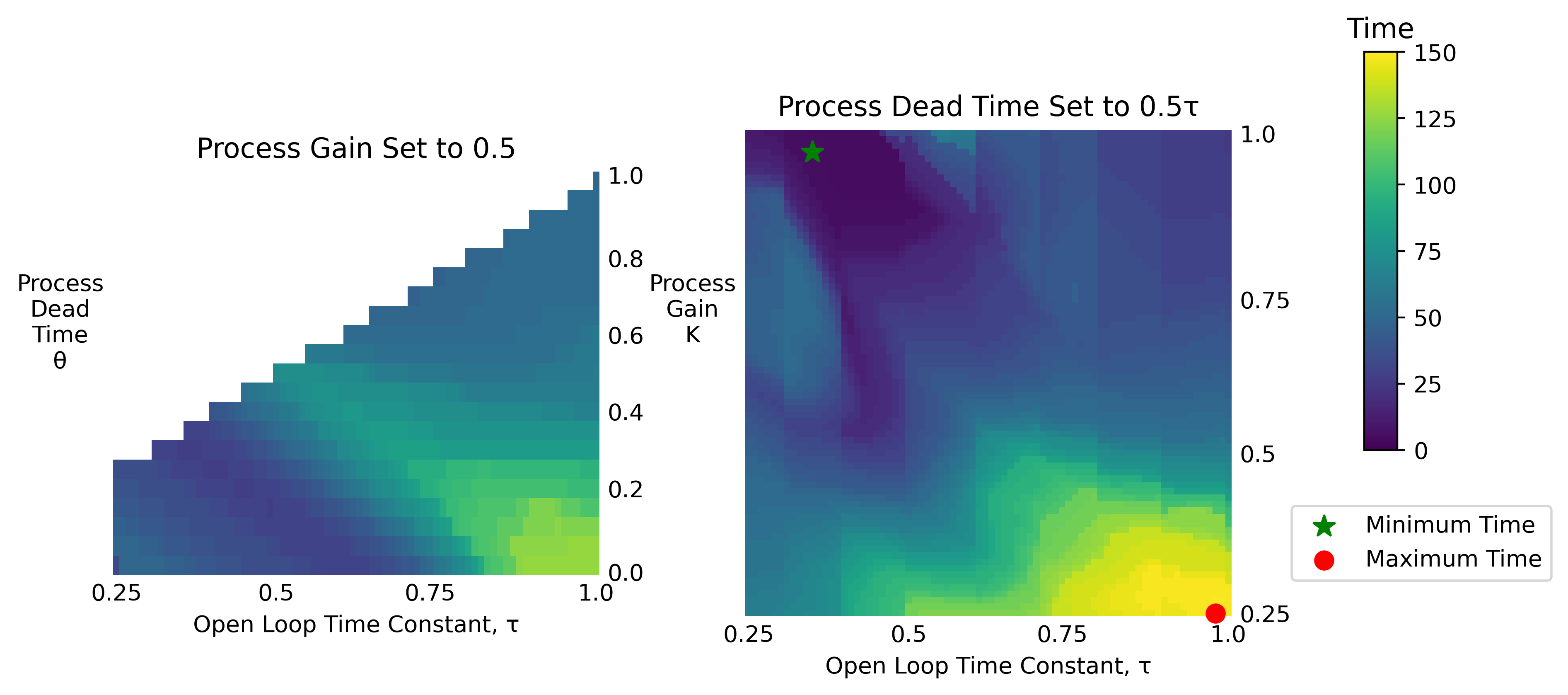}
    \caption{Online time required for both the $k_p$ and $k_i$ parameters to reach $\pm$10\% of their ultimate values.}
    \label{fig:time heatmap}
\end{figure}

\cref{subsec:asymptotic-performance} showed the asymptotic performance of the meta-RL PI tunings. Another important consideration is the online sample efficiency of the PI tuning; how fast do the controller parameters converge? \cref{fig:time heatmap} shows the time for both controller parameters to arrive within $10$\% of their ultimate values. The convergence of the tunings depends on the excitation in the system. In our experiments, excitation was created by setpoint changes every 11 units of time. The convergence speed could be increased with more excitation (or decreased with less). The meta-RL agent uses a sampling time of 2.75 units of time (i.e. the PI parameters are updated every 2.75 units of time; 4 times for each setpoint change).\par

Systems with large process gains and fast dynamics converge quickest, requiring just a single setpoint change (around 10 units of time). Systems with small gains and slow dynamics take longer to converge, requiring 13 setpoint changes to converge (around 140 units of time). \par

\begin{figure}
    \centering
    \includegraphics[width=\linewidth]{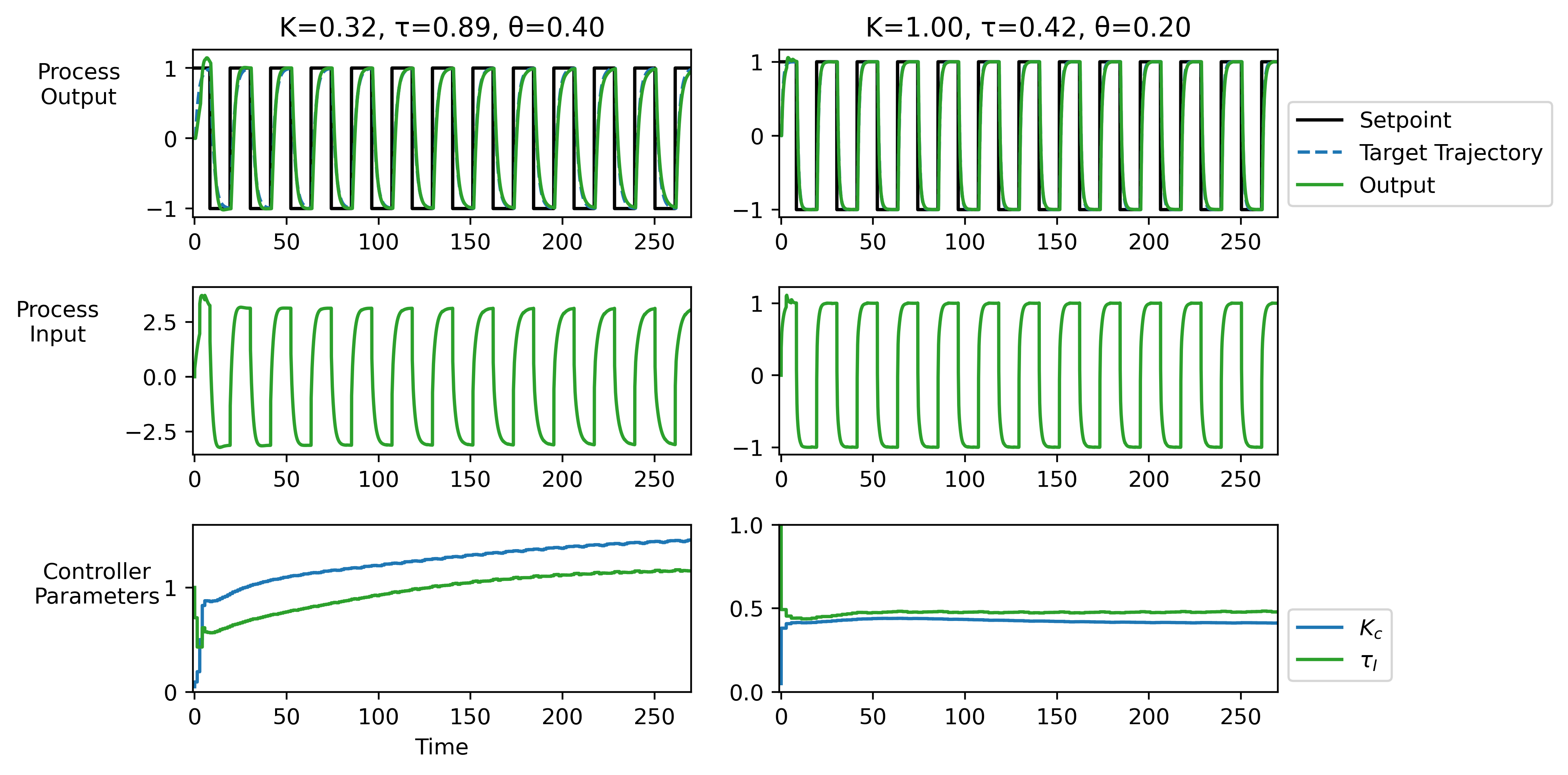}
    \caption{System output trajectories showing the convergence of the controller's PI parameters over time. The worst-case (left) and best-case (right) are selected from the heatmaps in \cref{fig:time heatmap}.}
    \label{fig:time trajectory}
\end{figure}

\cref{fig:time trajectory} shows the performance in the worst-case and best-case scenarios based on convergence time selected from \cref{fig:time heatmap}. Requiring over 13 setpoint changes to near convergence sounds undesirable, however from \cref{fig:time trajectory} we see even in this worst-case scenario, reasonable PI tunings are reached after a single setpoint change. The performance continues to improve with time to more closely match the target trajectory.\par

\subsection{Adaptive control using the meta-RL tuning algorithm}
\label{subsec:adaptation}
\begin{figure}
    \centering
    \includegraphics[width=\linewidth]{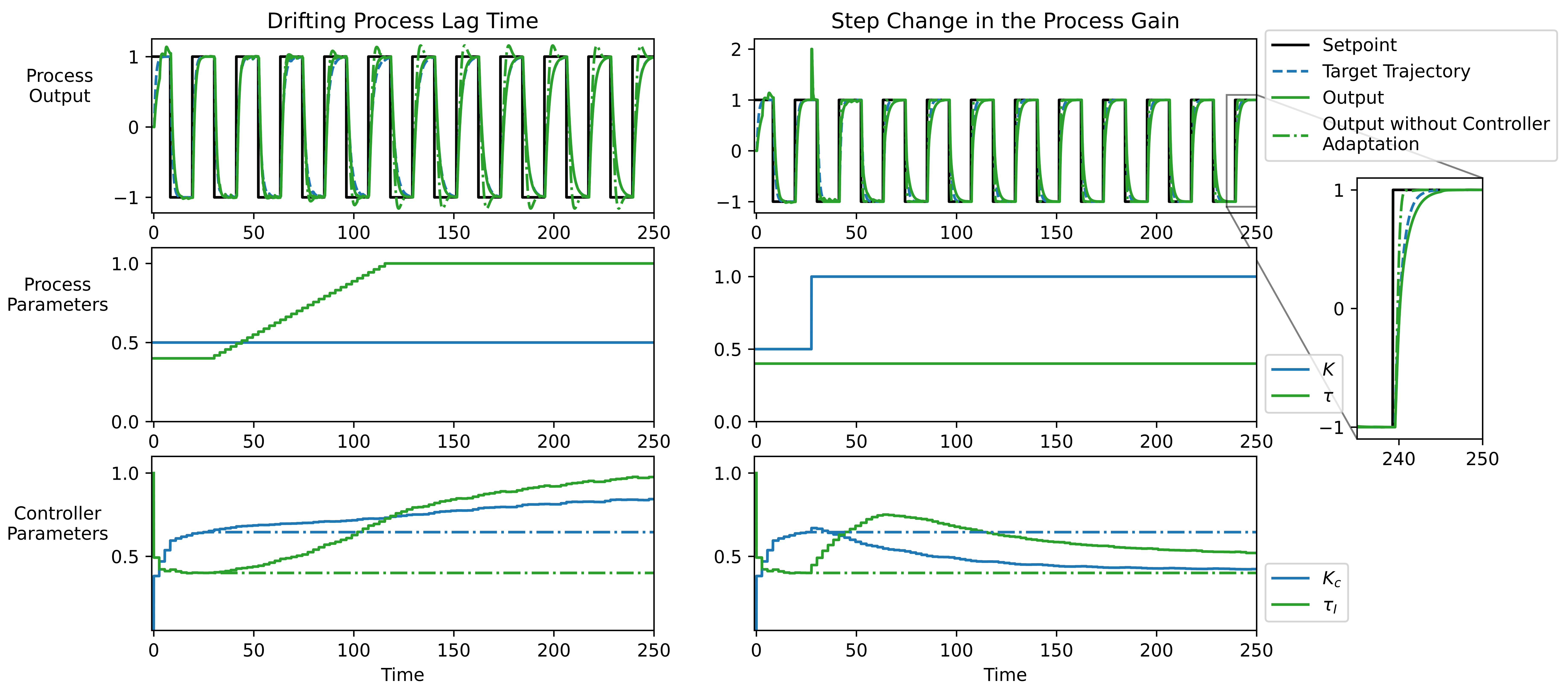}
    \caption{Meta-RL tuning with changing process dynamics. Solid lines show the controller parameters and process output with meta-RL in continuous operation. Fixing $K_c$ and $\tau_I$ at the the values initially produced by the meta-RL algorithm produces the dash-dot lines shown for comparison.}
    \label{fig:changing systems}
\end{figure}

In continuous operation, the proposed meta-RL tuning algorithm adapts effectively to changes in process dynamics. Such a change can be viewed as a move to a different location in the meta-RL agent's task distribution. Two sample scenarios involving significant changes to the process dynamics produce the results shown in \cref{fig:changing systems}. In the first, $\tau$ ramps up from 0.4 to 1.0; in the second $K$ steps up from $0.5$ to $1.0$. In both cases, a forgetting factor, $\gamma=0.99$, is applied to the meta-RL agent's hidden states at each time step. (This speeds up adaptation without noticeably affecting performance.) \cref{eq:rnn equation} can be modified to show how the forgetting factor is incorporated:
\begin{equation}
    z_t = f( \gamma W z_{t-1} + U x_t + b)
\end{equation}
The controller's parameters adapt to the changing system dynamics with very little disturbance to the system output (aside from an unavoidable disturbance when the process gain is suddenly doubled). In the case where the process time constant drifts, the meta-RL's adaptive tuning achieves a mean squared error of $0.006$ when tracking the target trajectory through a setpoint change---a $100$-fold improvement over the mean squared error of $0.0673$ when there is no meta-RL adaptation. For the step change in the process gain, the meta-RL adaptive tuning achieves a mean squared error of $0.0032$ while without adaptive tuning the mean squared error is $0.0290$ ($9$ times larger).\par

\subsection{Internal model representation}

The good simulation results in \cref{subsec:asymptotic-performance,subsec:time,subsec:adaptation}
suggest that information about the dynamics of the particular system being controlled
must somehow be embedded in method. 
To validate the hypothesis that the deep hidden states encode relevant information, we apply principal component analysis (PCA) to the ultimate deep hidden states of a trained model. We collect hidden state trajectories from simulations with different process gains and time constants but a constant ratio ${\tau}/{\theta}$. At the end of the simulations, the model has had time to converge to the final PI parameters and we expect the hidden states to differ primarily because of differences in the gain and time scale involved. Therefore, we expect differences between hidden states associated with different systems to be captured by two principal components (PCs) with very little loss of information.\par

\begin{figure}
    \centering
    \includegraphics[width=\linewidth]{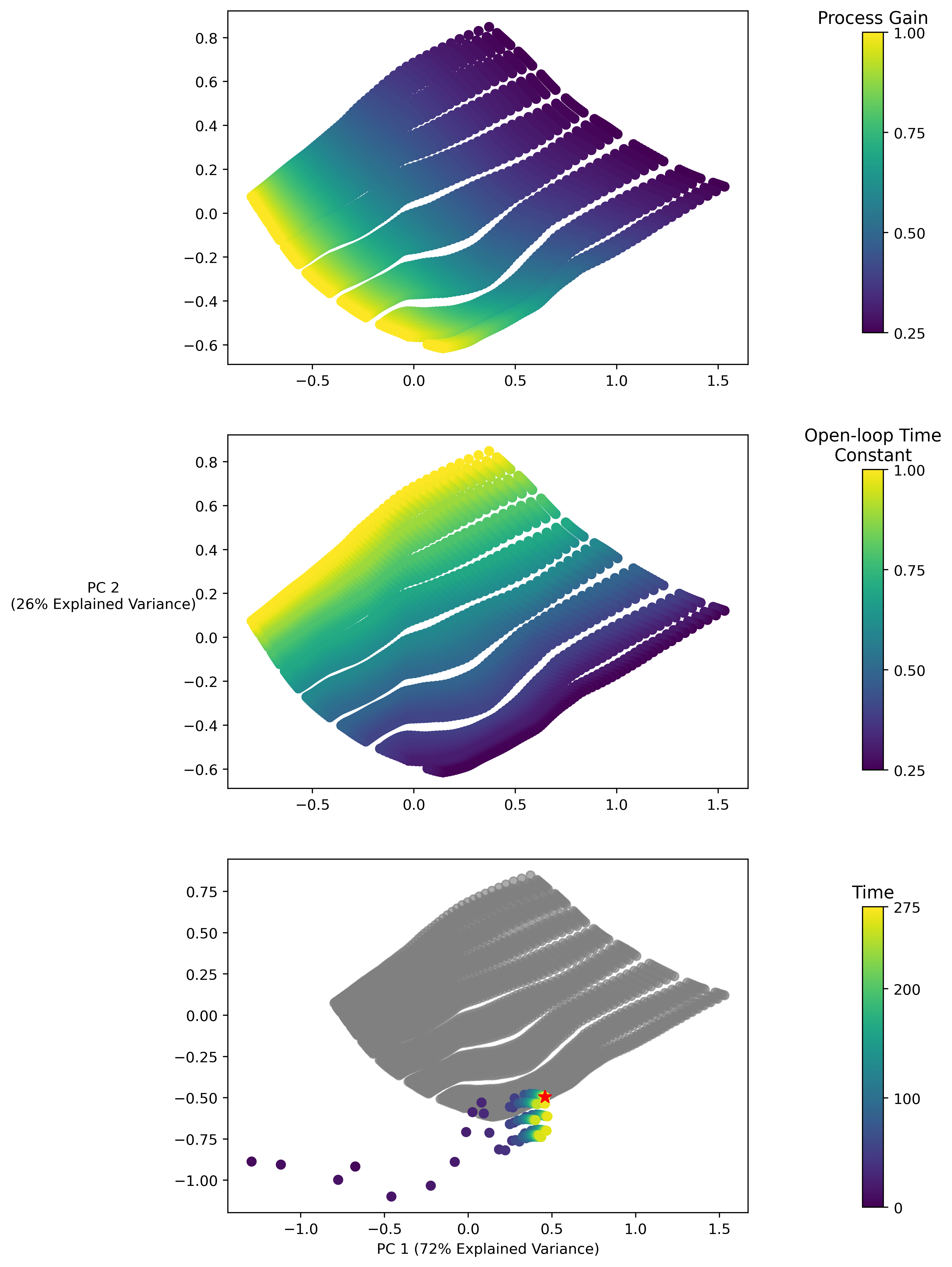}
    \caption{PCA projections of the ultimate deep hidden states from the meta-RL model after interacting with various systems. 
    (Process gains and time constants range from 0.25 to 1.0; the ratio of lag time to dead time is approximately constant.)
    The top two subplots color each projected point using the process gain and time constant of the underlying system, respectively. 
    The bottom subplot tracks the evolution of the meta-RL model's deep hidden state while interacting with the specific system where $K=0.75$, $\tau=0.25$, $\theta=0.20$.}
    \label{fig:pca}
\end{figure}

\cref{fig:pca} confirms this hypothesis. Two orthogonal components capture 98\% of the variance in the ultimate 100-dimensional deep hidden states. Projecting the hidden state into the plane spanned by these components, we show the process gain and time constant associated with each observation. The hidden states create a near-orthogonal grid based on these two parameters, whose variations act in complementary directions. Evidently the meta-RL model's hidden states constitute an internal representation of the process dynamics derived from closed-loop process data in a model-free manner.\par

The bottom subplot in \cref{fig:pca} shows how the deep hidden state evolves over time during a simulation involving a particular FOPTD system. The hidden states are initialized with zeros at the start of every episode. This corresponds to a point in lower left corner of the projected principal-component space, which the top subplot associates with systems having large process gains. This association is established during the process of training the meta-RL agent, and it admits a sensible interpretation. The system has ``learned'' to approach an unfamiliar process by assuming it has high gain. This leads to small control moves, which are appropriate until more information can be observed and incorporated into the controller design. The deep hidden state moves to a final point whose projection is highlighted in the figure: comparing the heatmaps in earlier subplots confirms that this point is associated with the correct values of $K=0.75$ and $\tau=0.25$.\par

\subsection{A simulated two-tank environment}
\label{subsec:tank-results}

The agent used above also performs well on a simulated version of the
practical two-tank control system detailed in~\citet{lawrence2022deep}.
Notable features of this example are the following:
\begin{itemize}
    \item The two-tank dynamics are nonlinear and slower than the FOPTD systems used for training the meta-RL agent. That is, the two-tank system is ``out of distribution''.
    \item The required operating regimes were not anticipated during training. Indeed, the meta-RL agent was only trained on step changes of $\pm 1$ starting from $0$.
\end{itemize}\par
We show how to apply the meta-RL agent to this novel environment despite these apparent obstacles. This simulated environment is a reasonable surrogate for a real apparatus: it is nonlinear, has a cascaded structure (for pump and flow control), and the pump, flow, level measurements are realizable through the use of filters. These dynamics are detailed in \citet{lawrence2022deep} but summarized here for completeness.\par

\subsubsection{Dynamics of the two-tank system}

\begin{singlespace}
\begin{table}[tbh]
\begin{center}
 \begin{tabularx}{\linewidth}{XXX}
\toprule
    Symbol &    Value or unit &    Description\\
\midrule
 $r_{\text{tank}}$ & 1.2065 (length) & Tank radius \\ 
 $r_{\text{pipe}}$ & 0.125 (length) & Outflow pipe radius \\ 
 $f_\text{max}$ & 80 (volume/time) & Maximum flow \\ 
 $f_c$ & 0.61 & Flow coefficient \\ 
 $\tau_{p, \text{in}, \text{out}, m}$ & 0.1, 0.1, 0.1, 0.2 (time) & Time constants \\ 
 $g$ & (length/$\text{time}^2$) & Gravitational constant \\ 
 $\ell$ & length & Tank level  \\ 
 $m$ & length & Filtered tank level \\ 
 $f_\text{in}$ & volume/time & Inflow  \\ 
 $f_\text{out}$ & volume/time & Outflow  \\ 
 $p$ & \% & Pump speed  \\ 
 $\bar{p}$ & \% & Desired pump speed  \\ 
\bottomrule
\end{tabularx}
\end{center}
\caption{Parameters and variables for the two-tank system. ``Tank'' here refers to the upper tank. The four time constants refer to the pump speed, inflow, outflow, and measured level, respectively. Length is in decimeters (dm), time is in minutes, volume is in liters. The tank height in our simulation is 12.192 dm.}
\label{table:tank}
\end{table}
\end{singlespace}

We consider the problem of controlling the liquid level in an upper tank, positioned vertically above a second tank that serves as a reservoir. Water drains from the tank into the reservoir through an outflow pipe, and is replenished by water from the reservoir delivered by a pump whose flow rate is our manipulated variable. More precisely, two PI controllers are in operation: For a desired level, one PI controller outputs the desired flow rate based on level tracking error. This flow rate is then used as a reference signal for the second PI controller, whose output is the pump speed. The first is referred to as the ``level controller'' and the second as the ``flow controller''. System parameters, values, and descriptions are given in \cref{table:tank}.\par

The system dynamics are based on Bernoulli's equation, $f_{\text{out}}\approx f_c\sqrt{2 g\ell}$, and the conservation of fluid volume in the upper tank:
\begin{equation}
\frac{d\hfil}{dt}\left(\pi r_{\text{tank}}^2\ell\right)
= \pi r_{\text{tank}}^{2} \dot\ell
= f_{\text{in}} - f_{\text{out}}.
\end{equation}
(We use dot notation to represent differentiation with respect to time.)

Our application involves four filtered signals, with time constants
$\tau_p$ for the pump,
$\tau_{\text{in}}$ for changes in the inflow,
$\tau_{\text{out}}$ for the outflow,  and
$\tau_m$ for the measured level dynamics. We therefore have the following system of differential equations describing the pump, flows, level, and measured level:
\begin{align}
\tau_p \dot{p} + p &= \bar{p}\label{eq:pump}\\
\tau_{\text{in}} \dot{f}_{\text{in}} + f_{\text{in}} &= f_{\text{max}} \left(\frac{p}{100}\right)\label{eq:fin}\\
\tau_{\text{out}} \dot{f}_{\text{out}}+ f_{\text{out}} &= \pi r_{\text{pipe}}^{2} f_{\text{c}} \sqrt{2 g \ell}\label{eq:fout}\\
\pi r_{\text{tank}}^{2} \dot{\ell} &= f_{\text{in}} -  f_{\text{out}}\label{eq:level}\\
\tau_m \dot{m} + m &= \ell. \label{eq:measure}
\end{align}
To track a desired level\footnote{Barred variables are used to denote setpoints. For example, $\bar{\ell}$ represents the tank level setpoint.}  $\bar{\ell}$, we can employ level and flow controllers by including the following equations:
\begin{align}
\bar{p} &= \text{PI}_{\text{flow}} (\bar{f}_{\text{in}} - f_{\text{in}})\label{eq:PIDflow}\\
\bar{f}_{\text{in}} &= \text{PI}_{\text{level}}(\bar{\ell} - m)\label{eq:PIDlevel}.
\end{align}
Equations (\ref{eq:PIDflow})--(\ref{eq:PIDlevel}) use shorthand for PI controllers taking the error signals $\bar{f}_{\text{in}} - f_{\text{in}}$ and $\bar{\ell} - m$, respectively. For our purposes, $\text{PI}_{\text{flow}}$ is fixed and a part of the environment, while $\text{PI}_{\text{level}}$ is the tunable controller.\par

This mathematical description is given to provide intuition for our control system. For the following results, we emphasize that the meta-RL agent was not trained on data from this environment, yet it iteratively fine-tunes the controller $\text{PI}_{\text{level}}$.\par

\begin{figure}
\begin{center}
    \includegraphics[width=0.75\linewidth]{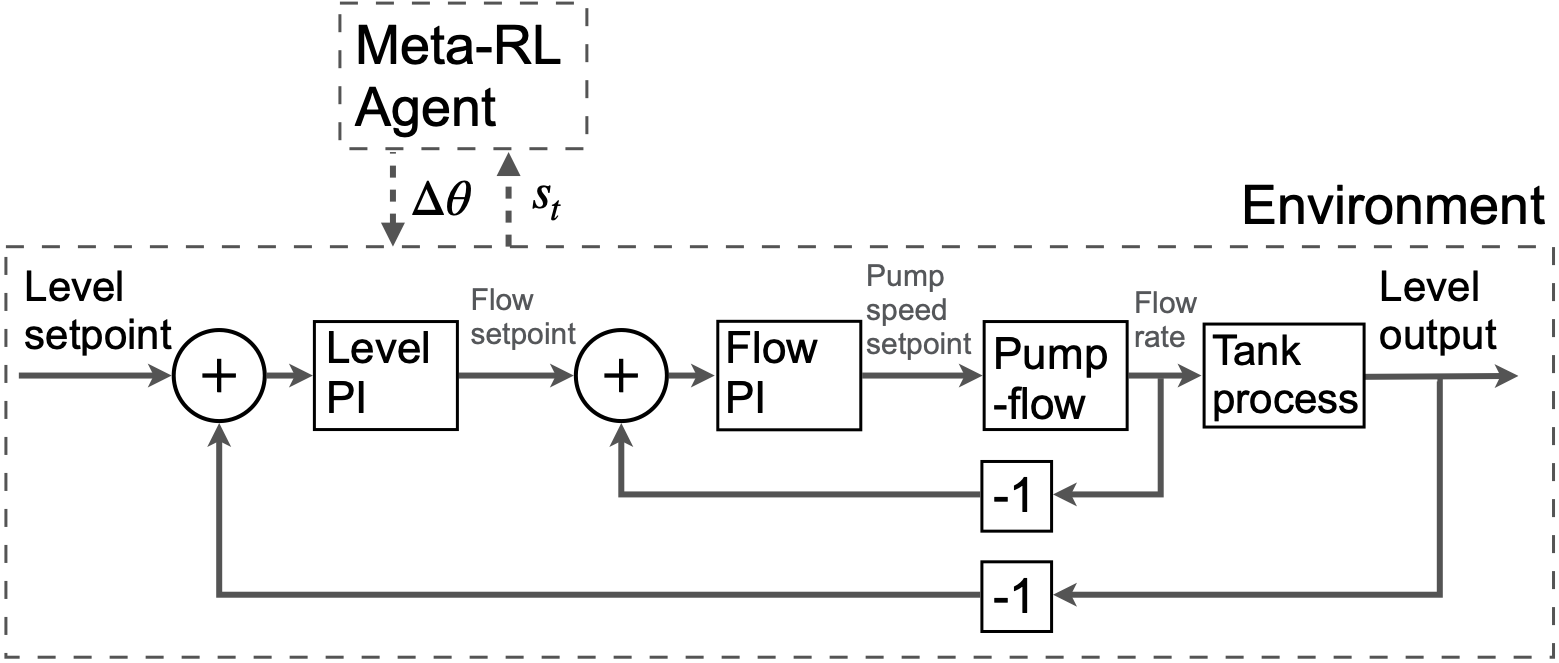}
    \caption{A schematic of the simulated nonlinear two-tank control system corresponding to \crefrange{eq:pump}{eq:PIDlevel}. The ``pump-flow'' block is modeled by \crefrange{eq:pump}{eq:fin}; The ``tank process'' is modeled by \crefrange{eq:fout}{eq:measure}; the level and flow controllers in \cref{eq:PIDflow,eq:PIDlevel} output flow setpoints and pump speed setpoints, respectively. The meta-RL agent generates incremental changes to the PI parameters of the level controller.}
    \label{fig:tank_schematic}
\end{center}
\end{figure}

\subsubsection{Adapting the meta-RL model to the two-tank system}

While the two-tank system is nonlinear, an accurate first order approximation of the dynamics relating the level in the tank to the pump flow rate setpoint is:

\begin{equation}
    G(s) = \frac{1.7}{55s+1}e^{-13s}
    \label{eq:tank_tf}
\end{equation}

For a realistic example of how the meta-RL tuning algorithm can be used, we assume only a crude approximation of the process' dynamics is available. The following crude model will be used to set up the meta-RL tuning algorithm:

\begin{equation}
    \widehat{G}(s) = \frac{1.2}{30s+1}
    \label{eq:bad tank tf}
\end{equation}
Crucially, the meta-RL agent is still interacting with the full two-tank system, given by \crefrange{eq:pump}{eq:PIDlevel}, and illustrated in \cref{fig:tank_schematic}. This model is simply used for data processing purposes, as we demonstrate below. \cref{eq:tank_tf} is only used to facilitate discussion about how a crude model compares to an accurate for meta-RL adaptation.\par
Our objective is to use the meta-RL algorithm to control the tank level around the operating region of 50--60 cm. First, we need to augment the process data to match the data distribution used to train the model (centered at 0, ranging from $-1$ to $1$). To do this, we first apply a constant control action to bring the tank level into the desired operating region ($u=12$ liter/min). Next, all process data has the mean (55 cm) subtracted and is scaled down by a factor of 10. This brings the data the meta-RL agent observes into alignment with its training distribution. Scaling the data also has the effect of decreasing the gain in the apparent process model (\ref{eq:bad tank tf}) to 0.12.\par

Next, we adjust the controller gain. The meta-RL algorithm is equipped to handle systems with process gains ranging from 0.25-1.0. By scaling the controller's output by $\frac{0.5}{0.12}$, we geometrically centre the model in \cref{eq:bad tank tf} to appear to the meta-RL agent as a system with $k=0.5$. If the estimated process gain used to set up the meta-RL agent is incorrect by any factor between $0.5\times$ to $2.0\times$, the true process gain will still fall within the task distribution. In this case, the true process gain of 1.7 appears as a process gain 0.71 to the meta-RL agent.\par

Next, we select an appropriate sampling time. By picking a slow sampling time, the tank's dynamics appear faster from the perspective of the meta-RL agent. To geometrically center the time constant in \cref{eq:bad tank tf} to the meta-RL's task distribution, we set the sampling time to every  $\frac{30}{0.5}=60$ seconds. The true time constant of 55 seconds then appears as a time constant of 0.92 to the meta-RL agent.\par

Through data augmentation, controller gain adjustment, and sampling time adjustment, the meta-RL agent's task distribution can be adapted to many ``out-of-distribution'' systems as long as the \emph{magnitudes} of each parameter can be estimated.\par 

Alternatively, if a meta-RL agent is being created for a particular application where there is a very coarse understanding of the process dynamics, the agent could be trained across a wide distribution of possible process dynamics to avoid the need for data augmentation and directly deploy the meta-RL agent on the system as in the previous examples. However, the advantage of direct deployment without data augmentation comes at the expense of training a meta-RL agent from scratch. Both these meta-RL approaches avoid the disadvantages of conventional RL methods: the need for very accurate estimates of the process dynamics or additional online fine-tuning to deal with plant-model mismatch.\par

\subsubsection{Results}

\begin{figure}
    \centering
    \includegraphics[width=\linewidth]{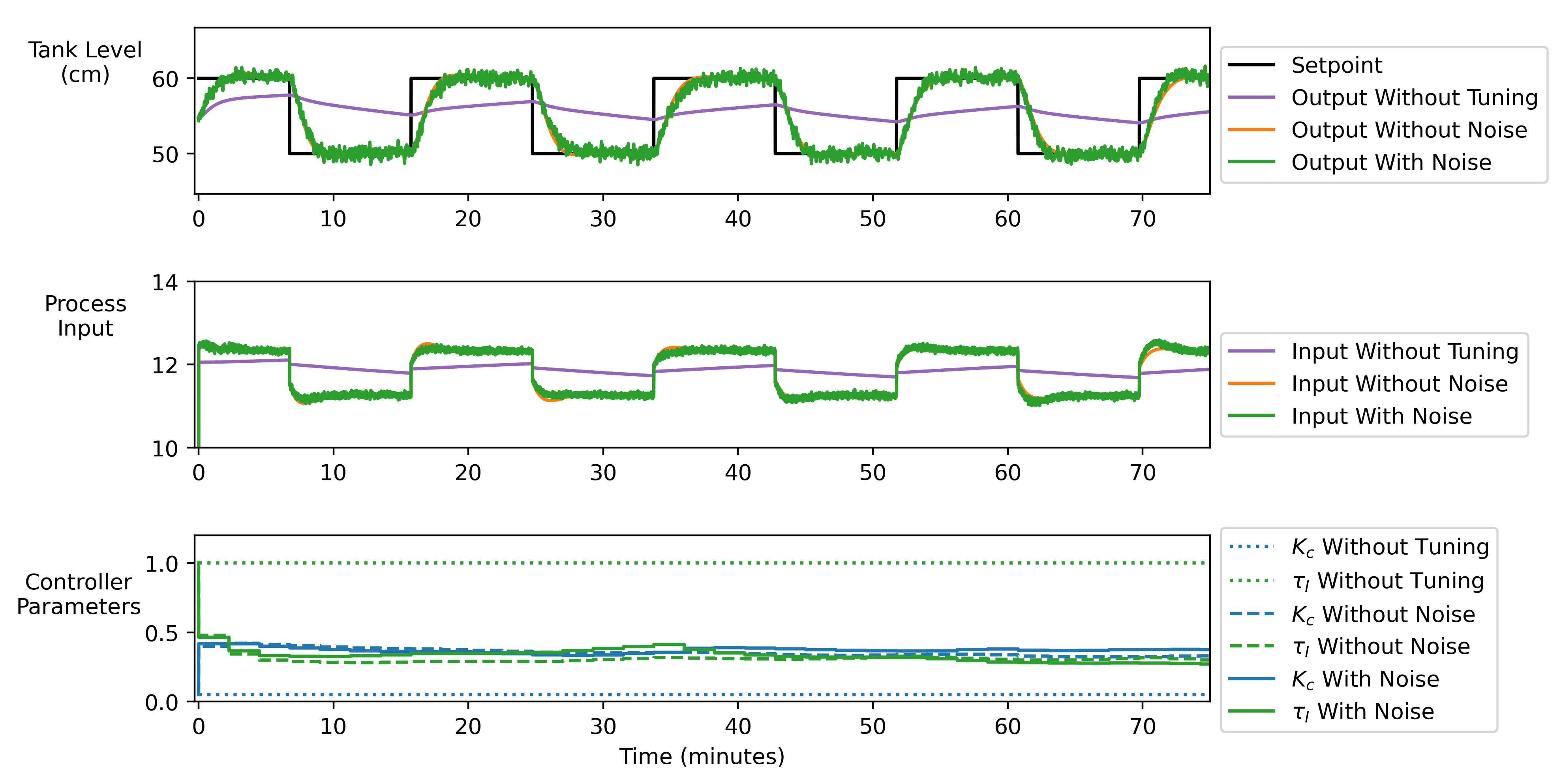}
    \caption{Performance of the meta-RL tuning algorithm for controlling the water level in a nonlinear two-tank system both with and without measurement noise. The performance without the meta-RL tuning is also shown as a point of reference.}
    \label{fig:tank}
\end{figure}

\cref{fig:tank} shows the tuning performance of the meta-RL agent on the two-tank system. After just one setpoint change, the meta-RL agent is able to find reasonable PI parameters for the system, demonstrating it is effective not just on true FOPTD systems, but also on nonlinear systems which can be approximated with FOPTD models. This example also contextualizes the sample efficiency of the meta-RL algorithm by providing an example with real units of time. For a system with a time constant around 1 minute and a dead time of around 13 seconds, it takes around 4 minutes for the PI parameters to nearly converge. \cref{fig:tank} also shows the performance of the meta-RL agent when there is noise of $\pm$1 cm added to the tank level measurements. Despite the meta-RL agent not being trained on systems with noise, we see the agent's performance is not significantly affected by this change. \par

This case study shows that the meta-RL algorithm can apply to a very large variety of processes. While a process model is not needed for the meta-RL algorithm to work, the \emph{magnitude} of the process gain and time constant must be known so the process data can be properly augmented. The task of scaling the gains and process dynamics needs to be automated for successful industry acceptance and this is an area for future work.\par

\section{Conclusion}
\label{sec:conclusion}

This work presents a meta-RL approach to tuning fixed-structure controllers in closed-loop without explicit system identification and demonstrates the approach using PI controllers. The method algorithm can be used to automate the initial tuning of controllers and, in continuous operation, to adaptively update controller parameters as process dynamics change over time. Assuming the magnitude of the process gain and time constant are known, the meta-RL tuning algorithm can be applied to any system which can be reasonably approximated as FOPTD\footnote{The present work focuses on FOPTD systems with $\tau>\theta$, however the results could be extended to dead-time dominant systems by expanding the task distribution $p_{\text{meta}}$.}. \par

A major challenge of applying RL to industrial process control is sample efficiency. The meta-RL model presented in this work addresses this problem by training a model to control a large distribution of possible systems offline in advance. The meta-RL model is then able to tune fixed-structure process controllers online with no process-specific training and no process model. There are two key design considerations which enable this performance. First is the inclusion of a hidden state in the RL agent, giving the meta-RL agent a memory it uses to learn internal representations of the process dynamics through process data. Second is constructing a value function which uses extra information in addition to the RL state. Conditioning the value function on this additional information, $V_{\phi}(s,\zeta)$ as opposed to $V_{\phi}(s)$, improves the training efficiency of the meta-RL model.\par

Industrial priorities suggest further investigation of the promising meta-RL framework presented here. For example, the training procedure should incorporate noisy process data and process disturbances of the sort often seen in real-world settings. The methods stability should also be investigated more deeply. (Using RL methods to produce PI parameters rather than to provide direct control inputs has the advantage of allowing access to known stability criteria. This is clearly relevant in practice, and may also suggest a feasible approach to future theoretical work.) The versatility of the meta-RL algorithm could also be improved by adding derivative action and extending the task distribution to incorporate a greater diversity of processes, including integrating processes and processes with higher order dynamics. Moreover, the task distribution could be extended to encompass both different process dynamics and different control objectives -- a complex process may require fast control for certain control loops and slower, smoother control for others. Finally, to add value to industry outside of continuous online tuning, we suggest exploring whether the meta-RL agent can be trained to identify when PID controllers should be retuned and what perturbation is needed for controller tuning (without relying on external sources of excitation, such as setpoint changes).

\section*{Declaration of competing interest}
\label{sec:Declaration}
The authors declare that they have no known competing financial interests or personal relationships that could have appeared to influence the work reported in this paper.

\section*{Acknowledgements}
\label{sec:acknowledgements}
We gratefully acknowledge the financial support of the Natural Sciences and Engineering Research Council of Canada (NSERC) and Honeywell Connected Plant.

\footnotesize
\bibliographystyle{elsarticle-num-names}
\bibliography{main.bib}

\appendix
\normalsize
\section{Meta-RL implementation details}

\captionsetup{labelformat=AppendixTables}
\setcounter{table}{0}
\setcounter{figure}{0}

\begin{table}[H]
    \begin{center}
    \caption{Hyperparameters used to train the meta-RL network.}
    \label{tab:network-hyperparameters}
    \begin{tabularx}{\linewidth}{XX}
      \toprule
      Hyperparameter & Value \\
      \midrule 
      Hidden layer size & 100 \\
      Recurrent cell type & GRU \\
      Activation function for feedforward layers & Leaky-ReLU\\
      Optimizer & Adam \\
      Initial learning rate & $3 \times 10^{-4}$\\
      Episode length & 40 steps (110 time units)\\
      Sequence length for backpropagation & 40 steps\\
      Training episodes per epoch & 300\\
      Epochs & 2500\\
      Discount factor$^{\star}$ & 0.99 \\
      GAE $\lambda^{\star}$ & 0.95\\
      Policy iterations$^{\star}$ & Up to 20\\
      Value iterations$^{\star}$ & 40\\
      Maximum KL divergence$^{\star}$ & 0.015\\
      Regularization penalty on $\Delta k_p$, $\beta_1$ & 0.5\\
      Regularization penalty on $\Delta k_i$, $\beta_2$ & 0.5\\
      \bottomrule 
    \end{tabularx}
    \caption*{$^{\star}$These hyperparameters are specific to PPO or RL more generally. The reader is referred to the original PPO paper by \citet{schulman2017proximal} for further explanation of these hyperparameters.}
    \end{center}
\end{table}

The key details of our meta-RL implementation are presented in \cref{tab:network-hyperparameters}.
Here we report the impacts of two other significant design decisions.

\subsection*{The role of ``privileged information'' during training}

In \cref{subsec:agent}, additional information outside of the RL state $s_t$ is used to train the value function. To investigate whether this additional information influenced the meta-RL agent's performance, an ablation study is conducted. The meta-RL agent's value function is trained with and without information about the process dynamics for an equal number of epochs. The performance of each meta-RL agent is presented in Figure A1. The meta-RL agent's performance is significantly better when the value function is trained with this additional information. The meta-RL agent's worst-case setpoint tracking error as measured by the mean squared error from the target trajectory for a step change from $-1$ to $1$ is $0.467$ without this information compared to $0.030$ with this information.

\begin{figure}
    \centering
    \includegraphics[width=\linewidth]{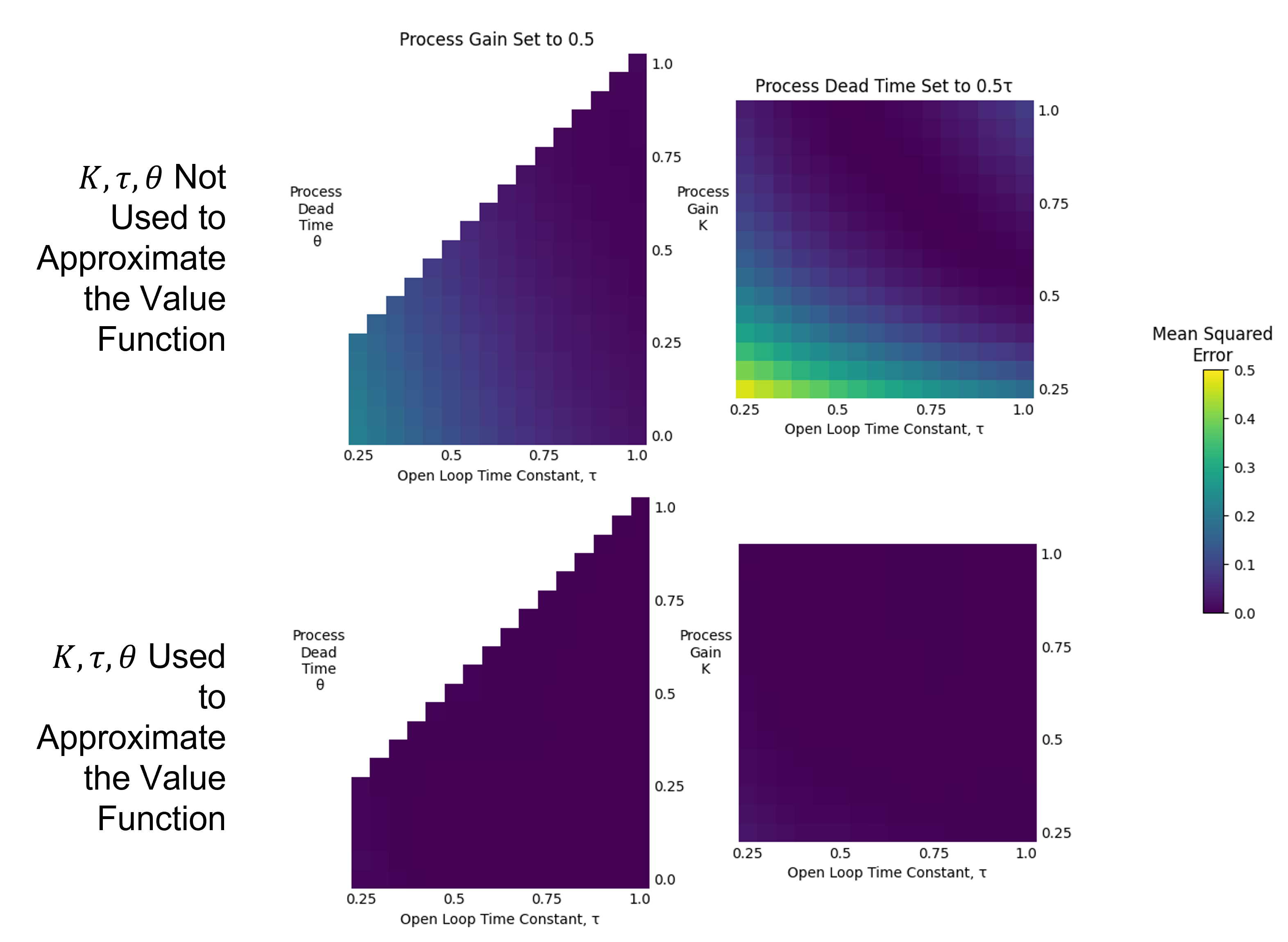}
    \caption{Performance of the meta-RL agent when the value function is trained without information about the process dynamics (top) vs. with this information (bottom).}
    \label{fig:appendixpic}
\end{figure}

\subsection*{The role of regularization in the reward function}

\begin{figure}
    \centering
    \includegraphics[width=\linewidth]{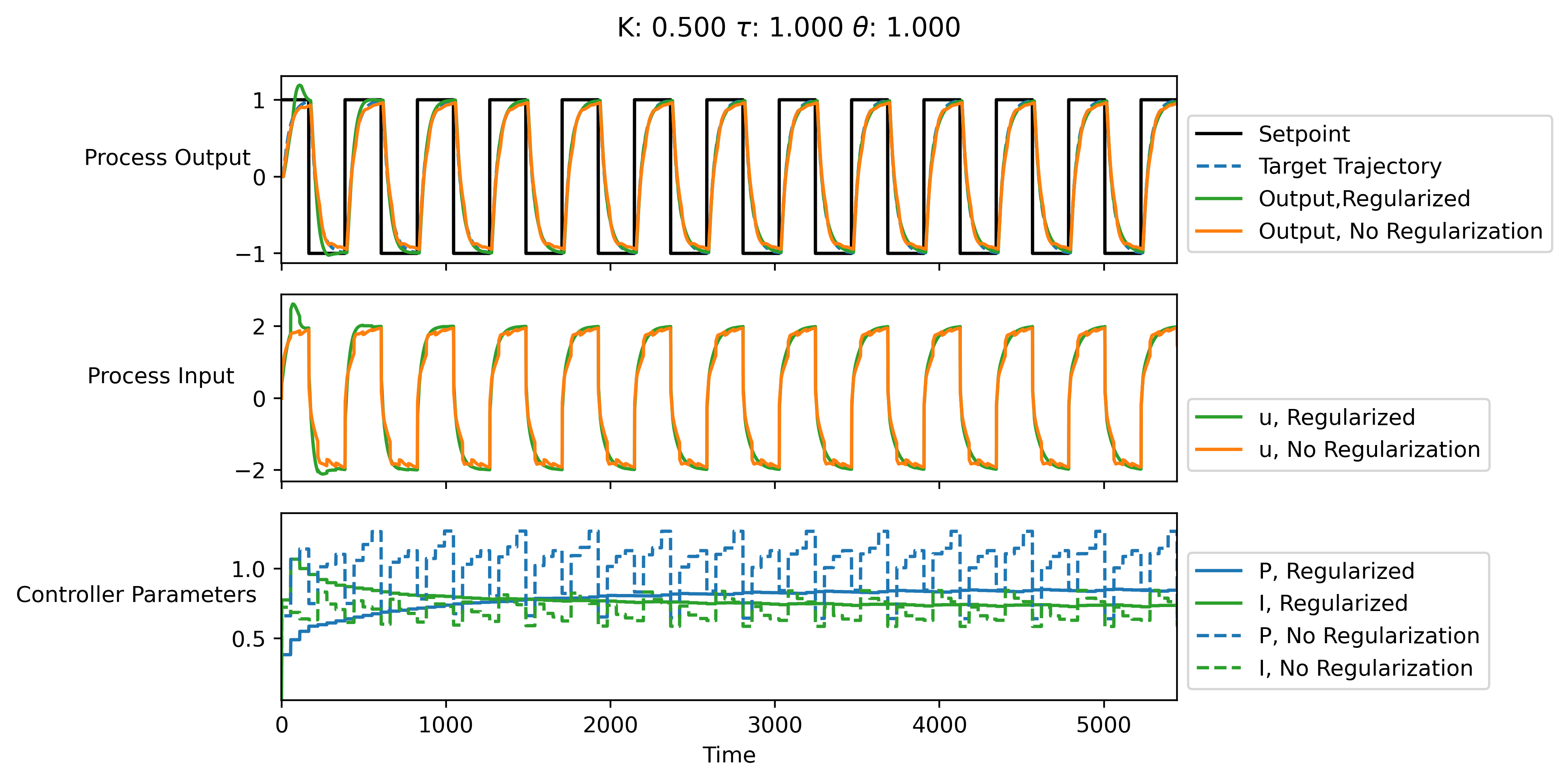}
    \caption{Performance of the meta-RL agent with and without cost regularization.
    For regularized trajectories, $\beta_1=\beta_2=0.5$ in \cref{eq:cost};
    trajectories with no regularization have $\beta_1=\beta_2=0$. The system has $K=0.5$, $\tau=1.0$, $\theta=1.0$.}
    \label{fig:appendixlastpic}
\end{figure}

The cost in \cref{eq:cost} includes penalty terms proportional to the size of the parameter updates proposed by the meta-RL agent. These regularization terms aid in the convergence of the PI parameters. Sample trajectories produced by the meta-RL agent trained with and without this regularization are shown for comparison in \cref{fig:appendixlastpic}.

\end{document}